# Transparent Hybrid Anapole metasurfaces with negligible electromagnetic coupling for phase engineering

Alexey V. Kuznetsov[1], Adrià Canós Valero[1], Mikhail Tarkhov[4], Vjaceslavs Bobrovs[2], Dmitrii Redka[2,3], and Alexander S. Shalin[1,2,5]

[1]ITMO University, Kronverksky prospect 49, 197101, St. Petersburg, Russia

[2]Riga Technical University, Institute of Telecommunications, Latvia, Riga, Azenes street 12, post code 1048

[3]Electrotechnical University "LETI" (ETU), 5 Prof. Popova Street, Saint Petersburg 197376, Russia

[4]Institute of Nanotechnology of Microelectronics of the Russian Academy of Sciences (INME RAS), Moscow, Nagatinskaya street, house 16A, building 11

[5]Kotel'nikov Institute of Radio Engineering and Electronics of Russian Academy of Sciences (Ulyanovsk branch), Goncharova Str.48, Ulyanovsk, Russia, 432000

**Abstract:** All-dielectric nanophotonics has become one of the most active fields of research in modern optics, largely due to the opportunities offered by the simultaneous resonant control of electric and magnetic components of light at the nanoscale. In this rapidly evolving scenario, the possibility to design artificial Huygens sources by overlapping electric and magnetic resonances have established a new paradigm in flat optics, bringing devices closer to efficient wavefront shaping with a direct phase engineering at the level of the individual meta-atoms. However, their efficiency is fundamentally limited by the near field coupling between the constituents of the metalattice. In this work, we challenge this well-conceived notion and propose an alternative concept to achieve phase control and full transmission in metasurfaces, based on the unusual properties of the nonradiating sources known as Hybrid Anapoles (HAs). We analyze theoretically an array of such sources and demonstrate that HAs are characterized by a negligible coupling with their neighbors. Therefore, in contrast to Huygens particles, the proposed sources can operate as individual meta-atoms even in highly compact designs, becoming robust against strong disorder and preserving its characteristics when deposited on dielectric substrates. Remarkably, the phase of the transmitted wave can be modulated with negligible reflection. To illustrate the capabilities of our platform, we also utilize a disordered HA array to implement a controlled phase modulation to an ultrafast gaussian pulse. The results of our study represent a departure from the currently established designs and open an avenue towards the realization of new devices for flat optics with unprecedented efficiency.

**Keywords:** all-dielectric metasurfaces, hybrid anapoles, nonradiating sources, resonant nanoantennas, flat optics

## 1 Introduction

Over the past few years, all-dielectric nanophotonics has become one of the most active fields in optics [1–4]. High-index subwavelength nanostructures represent a radical departure from the field of plasmonics, paving the way towards efficient control of light at the nanoscale in the absence of dissipation losses, and offering CMOS compatibility. Careful shape and dispersion engineering of subwavelength dielectric cavities allows to excite and tune both electric and magnetic resonances [5–7]. Optical structures composed of such artificial elements, commonly termed 'meta-atoms', feature a plethora of exotic effects not accessible with natural materials still being actively explored up to date, such as artificial magnetism[8], nonradiating sources, supercavity modes and bound states in the continuum, efficient second and third harmonic generation [9], or spin-orbit conversion[10].

Among all the promising phenomena uncovered in this versatile research platform, directional Huygens sources, caused by the simultaneous overlap of electric and magnetic resonances with opposite parity [11,12], hold enormous interest in the flat optics community, since two-dimensional subwavelength arrays -metasurfaces- of dielectric Huygens nanodisks with different size can imprint varying phases to an incident field while featuring unity transmission [13], giving birth to the vibrant field of all-dielectric 'meta-optics'. As a result of these developments, a number of applications have already been devised, such as ultrathin metalenses [14,15], dynamic control of transmission [16], anomalous refraction [17], beam steering [18], and holograms [19–21].

To achieve the ultimate goal of spatial wavefront control with subwavelength resolution, Huygens metasurfaces are required to overcome fundamental bottlenecks [22], usually ascribed to inter-element coupling [17,23]. When placed on an array, there exists an important cross-talk between Huygens meta-atoms even for separations in the order of the wavelength [24], complicating the engineering of phase at the level of the individual constituents. Several attempts have been made to minimize this issue; for instance, in Ref. [17] the authors propose the use of carefully optimized supercells, while in Ref. [25] the effect of coupling is exploited precisely in order to imprint varying phases to distinct regions of an array. These strategies have allowed to push the performance limits of Huygens metasurfaces, but the question remains whether a direct mapping of phase to the individual meta-atom will ever be reachable. Large-scale numerical optimization of Huygens metasurfaces has indeed been shown to improve their performance, but in no means gradient-based approaches allow to minimize interparticle interactions, rather they maximize the efficiency of the ensemble [26]. As a result, the geometrical degrees of freedom must be carefully adjusted at each stage of the design, and special care must be taken to include the influence of period, underlying substrate, and disorder in the array. Particularly the latter can drastically modify the expected transmitted phase with respect to the periodic metasurface [27,28]. Because of the aforementioned reasons, the efficiency of such devices might be fundamentally limited[22].

Recently, an alternative to the Huygens effect allowing to realize unity transmission in a metasurface has been proposed. Specifically, the so-called 'transverse kerker' effect, characterized by out-of-phase dipoles and quadrupoles in a scatterer, was shown to lead to an 'invisible' metasurface, where light traversed the array without any perturbation [29]. Thus, the rich multipolar toolbox offered by all-dielectric nanostructures still hosts a number of surprising effects that can be exploited to improve the performance of meta-devices, beyond Huygens.

In this work, we propose a novel mechanism to achieve unity transmission and phase control based on the physics of a recently observed nonradiating state, the HA [30]. Nowadays, the so-called anapole states, originating from the destructive interference of the electric and toroidal dipole moments [31], have already revolutionized nanophotonics, demonstrating enhanced second and third harmonic generation [32,33], giant photothermal nonlinearities [34] and 'dark' lasers [35]. HAs constitute the 'new' generation of anapoles, arising due to the simultaneous destructive interference of all dominant multipole moments with their toroidal counterparts [30]. While the possibility to overlap anapoles was first discussed in [36], only recently HAs were theoretically proposed and experimentally confirmed to occur in dielectric nanocylinders by inducing a degeneracy of two high order modes with different symmetry[30]. Unlike conventional anapoles, they have a mixed electric and magnetic nature, in some cases displaying helicity singularities [34]. Furthermore, they have been shown to outperform their counterparts both in terms of scattering suppression and stored electromagnetic energy, and retain their nonradiating character in the presence of an arbitrary dielectric substrate [30].

Here, we investigate in detail the properties of HA-based metasurfaces. We first show analytically and numerically that their ultra-weak interaction with the environment naturally leads to unity transmission. However, relaxing the constrain for complete suppression of scattering, the multiresonant nature of HAs allows one to vary the transmitted phase within a wide spectral and geometrical range. Most importantly, unlike Huygens sources, inter-particle coupling is almost negligible. Thus, the HA particle approaches the ideal of a 'true' meta-atom. We showcase this ability designing highly compact silicon arrays with inter-particle wall-to-wall separations of 1/8$^{th}$ the incident wavelength in the visible range, as well as disordered HA-metasurfaces exhibiting an identical behavior to their periodic counterparts. We study the influence of a substrate and demonstrate the preservation of the transmission window when the metasurface is deposited over a broad range of dielectric materials, potentially facilitating their on-chip integration. Finally, as a proof-of-concept application, we modulate the phase of an ultrafast gaussian pulse transmitted with unity efficiency from a highly disordered HA metasurface deposited on a glass substrate, solely with the knowledge of the optical response of the periodic array.

## 2 Limitations of the Huygens meta-atom

The transmission characteristics of an arbitrary metasurface composed of identical resonators can be well described with the knowledge of the effective multipoles induced on a single meta-atom, taking into account their mutual interaction [37]. Then, the complex transmission coefficient $t(\omega)$ is a result of the linear addition of the complex multipole coefficients together with the direct transmission:

$$t(\omega) = 1 + \sum_i \left[ t_i^e(\omega) + t_i^m(\omega) \right], \quad (1)$$

where the superscripts $e, m$ stem for 'electric' and 'magnetic', and the summation runs over all the multipoles having a non-negligible effect on the scattering characteristics of the isolated meta-atom ($i=1$ dipoles, $i=2$ quadrupoles, etc). An expression similar to Eq.(1) can be obtained for the reflection coefficient $r(\omega)$. As we prove rigorously in the Supplementary Material S1, Eq.(1) is valid for all metasurfaces with subwavelength interparticle separations if a sufficient set of terms in the expansion are taken into account.

When the metasurface is constituted with particles much smaller than the wavelength, (i.e. $kr_{eff} \ll 1$, $r_{eff}$ being a characteristic dimension of the particle) only the dipoles play an important role [38–40]. For clarity, from now on we omit all arguments in the expressions, and refer the interested reader to Ref. [41] and the Supplementary Material S1 for details. Considering an x-polarized plane wave propagating in the -z direction with amplitude $E_0$, and assuming the metasurface is suspended in free space, Eq. (1) takes the simplified form:

$$t = 1 + \frac{C}{S_l}\left[ \tilde{p}_x + \frac{1}{c}\tilde{m}_y \right], \quad (2)$$

Where $C = ik/(2D_0)$, with $D_0 = \varepsilon_0 E_0$, and $S_l$ is the surface of the lattice unit cell. Similarly, the reflection coefficient is

$$r = \frac{C}{S_l}\left[ \tilde{p}_x - \frac{1}{c}\tilde{m}_y \right]. \quad (3)$$

The minus sign in Eq.(3) appears due to the opposite parity of the electric and magnetic dipoles, and is the key-enabling feature of Huygens metasurfaces. It gives rise to the well-known forward Kerker condition [11], implying that if the electric $\tilde{p}_x$ and magnetic $\tilde{m}_y$ dipoles are in-phase and fulfill the relation $\tilde{p}_x = \tilde{m}_y/c$, there is no reflection and therefore the transmitted amplitude $T = |t|^2 = 1 - |r|^2 = 1$ in the absence of absorption losses. Importantly, the transmitted phase can be varied with unity efficiency, as long as the phase difference between the multipoles is kept approximately constant.

One of the most attractive traits of Huygens metasurfaces is the notion of "meta-atom": a subwavelength unit cell with individually tailored optical response, in our case designed to provide a specific phase when incorporated into an array with dissimilar characteristics, for the implementation of different types of phase gradients. In reality, it has been shown that the meta-atoms are in general highly sensitive to their surrounding [24], and therefore additional and resource-consuming optimization steps, taking into account the metasurface as a whole, are required to reach the desired efficiency [24,25]. Unfortunately, during the process, the ideal notion of 'meta-atom' is lost [22], since one can no longer trace a one-to-one mapping between the response of the isolated particles and their response in the array.

The most important bottleneck is the mutual coupling between the particles in the array [24,42]. Indeed, the effective multipoles $\tilde{\mathbf{b}} = [\tilde{p}_x, \tilde{m}_y]^T$ entering Eqs.(2)-(3) can only be obtained from the isolated particle response $\mathbf{b} = [p_x, m_y]^T$ after including the influence of the scattered fields from all the particles in the array. By means of multiple scattering theory [41], one can derive a system of linear equations:

$$\tilde{\mathbf{b}} = (I - \mathcal{AS})^{-1} \cdot \mathbf{b}, \quad (4)$$

where two important physical terms can be distinguished; the *polarizability matrix* $\mathcal{A}$ is a diagonal matrix containing the electric and magnetic multipole polarizabilities of the isolated particle. The *coupling matrix* $\mathcal{S}$ quantifies the strength of the multipolar fields from each particle in the array (and its derivatives in the case of quadrupolar contributions and higher), evaluated at the center of mass of the particle (see Supplementary Material S1 and Ref.[41]). The elements of $\mathcal{S}$ are solely a function of the electromagnetic propagator in the medium evaluated at the positions of each constituent.

Customarily, in the analysis of Huygens metasurfaces, the dipolar polarizabilities entering $\mathcal{A}$ are assumed to follow a Lorentzian dispersion [13]:

$$\alpha_{p,m}(\omega) = \frac{d_0}{\omega_0^2 - \omega^2 - i\omega\gamma}. \tag{5}$$

This formula corresponds to a phenomenological damped harmonic oscillator with amplitude $d_0$ resonating at $\omega_0$, driven by an external field with angular frequency $\omega$. Here we encounter an important limitation of the Huygens meta-atom; indeed, near resonance, $\alpha_{p,m}(\omega_0)$ is drastically enhanced, and therefore *so is the coupling* induced by the product $\mathcal{AS}$ in Eq.(4). In order to minimize coupling, a general design rule states that the period of an array $l$ should be larger than $\sqrt{\sigma_{ext}}$, ($\sigma_{ext}$ the extinction cross section of the single particle), imposing clear limitations on their compactness and size of the unit cell, and complicating the design of applications. Moreover, when the metasurface is placed over a substrate, the effective multipoles must also take into account the scattered fields of their 'mirror' images [43]. Huygens metasurfaces, and in general any dielectric or plasmonic metasurface, are in consequence strongly influenced by the substrate.

## 3 Phase-Changing HA Metasurfaces

Clearly, an unconventional approach is needed to overcome the abovementioned difficulties. If we require a *true* meta-atom, we are bound to only exploit single particle effects, i.e., we can solely act on $\mathcal{A}$. A possible solution would then be to set $\mathcal{A} = 0$. Anapoles are promising candidates since they allow to resonantly suppress the electric dipole polarizability beyond the Rayleigh limit [31]. However, only HAs can simultaneously cancel all the dominant electric and *magnetic* multipole moments, even leading to complete transparency [30]. Namely, if we write $t$ up to the quadrupolar terms, we obtain

$$t = 1 + \frac{C}{S_l}\left(\tilde{p}_x + \frac{1}{c}\tilde{m}_y - \frac{ik}{6}\tilde{Q}_{xz} - \frac{ik}{2c}\tilde{M}_{yz}\right). \tag{6}$$

$\tilde{Q}_{xz}$ and $\tilde{M}_{yz}$ in Eq.(6) are, respectively, the excited components of the electric and magnetic quadrupolar tensors. At the anapole in an isolated particle, the exact electric dipole is suppressed by interference of the quasistatic electric dipole $p_x^0$ with the electric toroidal dipole $T_x^p$ in the now classical formula $p_x = p_x^0 + ikT_x^p = 0$. Conversely, at the HA all the dominant multipoles are resonantly suppressed via the interference with their toroidal counterparts [30]. From Eq.(6), the latter implies near unity transmission. This apparently simple realization lays the foundation of our novel proposal, *nonradiating* HA metasurfaces enabling full transmission. We must also point out that destructive interference occurs strictly outside the smallest spherical region enclosing the scatterer [44]. However, similarly to a conventional anapole, the fields within the particle remain strongly enhanced [30,31], but are much more confined in the high-index region due to the larger quality factors of the modes involved. As will be shown later, the combination of the two effects results in a negligible electromagnetic interaction in the near and far field regions.

In the following subsections, we carry out a detailed investigation of both periodic and aperiodic (disordered) arrays of HA meta-particles under normally incident plane wave illumination, as schematically depicted in Figure 1(a). We demonstrate how the highly sought properties of the HA are directly inherited by the metasurface, allowing for a resonant suppression of reflection, invariance with period, and robust protection from disorder, a unique trait of our design. Remarkably, we reveal that the transmitted phase can be varied within the region of strong scattering cancellation. We dedicate a specialized subsection to understand the influence of a dielectric substrate, demonstrating that the transmission band is preserved for a broad range of materials. In particular, we show that a glass substrate allows to double the available phase range. Altogether, our non-Huygens metasurfaces can therefore represent a new cornerstone in Mie optics.

## 3.1 Near unity transmission and phase control

First, we investigate periodic square arrays of HA meta-atoms similar to the ones proposed in Ref. [30], constituted of amorphous silicon, where we have also included dissipation losses (the full dispersion is provided in section S4 of the Supplementary Material). Figure 1(b) displays the calculated transmission (T) spectra for a HA metasurface with separation between walls of resonators $s = 300\,\text{nm}$. $\lambda_0$ indicates the wavelength featuring an almost complete overlap of the dipole and quadrupole anapoles in a single nanoparticle (refer to the Supplementary Material S5). Confirming the theory, a broad transmission band can be observed in the vicinity of $\lambda_0$. We note that unity transmission can be achieved if dissipation losses are neglected.

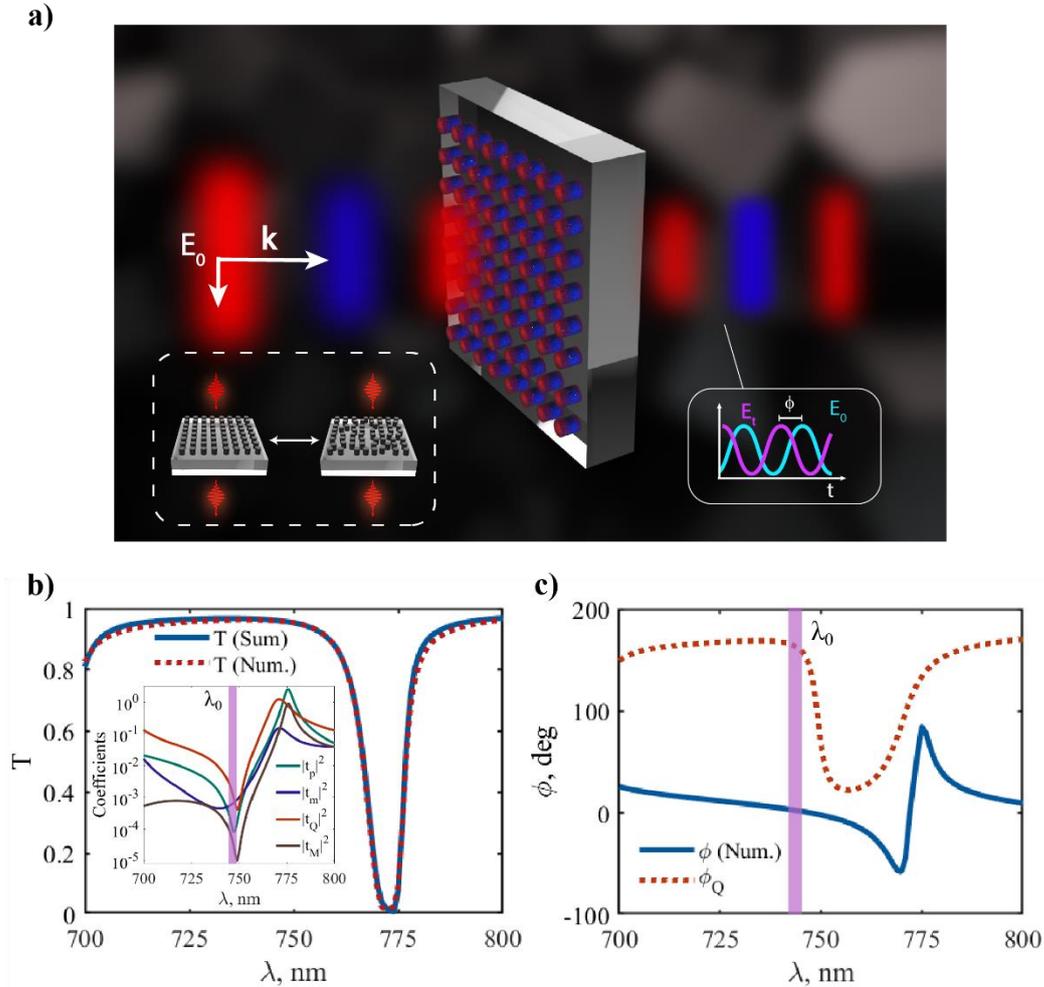

**Figure 1.** (a) Schematic overview of the considered Si metasurfaces composed of HA nanoparticles, illuminated with a normally incident plane wave $E_0$, illustrating its new functionalities: full transmission, phase control (right inset) and negligible electromagnetic coupling (right inset), allowing for the design of ultracompact, as well as disordered arrays retaining the single particle response. (b) Numerical and semianalytical [Eq.(6)] transmission spectra obtained for a HA metasurface with $s = 300$ nm (inset: absolute values of the dominant multipolar contributions). The geometrical

parameters of the meta-atom are: height 370 nm, and radius 130 nm. (c) Total transmitted phase $\phi$ obtained numerically, and phase variation experienced by the electric quadrupole $\phi_Q$.

The transmitted phase is exactly 0 at $\lambda_0$ [Figure 1(c)], and experiences a gradual increase along the transmission band of approximately 40 degrees. To gain insight into the physical origin of the phase change, we decompose the complex transmission coefficient with the help of Eq. (6) and the exact expressions for the multipole moments in Cartesian coordinates [45] (see Supplementary Material S3) . We stress that, unlike the scattering cross section, T is not the result of the direct sum of the absolute values of each individual multipole since they interfere. Instead, T is given by the squared modulus of Eq.(6), in full agreement with the numerical calculations [Figure 1(b)]. We first center our attention in the absolute values of the multipolar contributions to transmission [Inset of Figure 1(b)]. In direct analogy with the single particle behavior, we observe a resonant cancellation of all the multipole moments at $\lambda_0$, in the form of dips that extend along the whole transmission band. In the Supplementary Material S6 we demonstrate that their origin is indeed due to the interference of the quasistatic moments with their toroidal counterparts.

Complete destructive interference is achieved only at $\lambda_0$ (leading to a zero-phase shift). However, the multipolar fields are still strongly suppressed in its spectral vicinity. We stress that despite its small overall contribution, the electric quadrupole moment is the dominant multipole in the spectral range of interest, exceeding the electric dipole by an order of magnitude. The latter confirms that the near unity transmission *is not due to a generalized Kerker effect*[46] between the electric dipole and the electric quadrupole moments, but solely due to the resonant scattering suppression of the HA. Importantly, the quadrupole increases in an almost linear fashion towards shorter wavelengths, while its phase remains almost constant [Figure 1(c)]. Within the spectral region covered by the HA, despite its linear growth, the quadrupole is much smaller than unity, and thus displays negligible reflection. Taking these observations into account, the evolution of the transmitted phase can now be understood as due to the dephasing in the forward direction of the electric quadrupole field with the incident plane wave.

We now assess one of the most distinct advantages of our novel non-Huygens metasurfaces, i.e., their robustness against changes in inter-particle separation $s$ (Figure 2). As a comparison, we also included calculations with a less efficient non-Huygens metasurface being constituted solely of nanoparticles supporting the electric anapole [47]. Figure 2(a) shows the variation in the overall transmission as a function of period, evaluated at $\lambda_0$ (the minimum of the electric anapole has also been designed to occur at $\lambda_0$). Due to the small overall polarizabilities, both cases studied preserve unity transmission up to very small periods. HA metasurfaces, however, can successfully suppress reflection even at dramatically small separations between the constituents, i.e. in ultra-compact arrays with wall-to-wall distances reaching less than 6% the size of the incident wavelength. Furthermore, the transmitted phase is also virtually unaffected. In view of this result, we anticipate the possibility to realize ultrasmall pixels beyond what has been achieved so far with other platforms, e.g. for holography applications.

Such compact arrays cannot be analytically described with multiple scattering theory, i.e. Eq.(4) is no longer valid, since the multipole moments are not orthogonal and do not form a complete basis once neighboring particles cannot be enclosed by non-intersecting spherical surfaces[48]. However, as we proved analytically in the Supplementary Material S1, the multipole decomposition of reflection and transmission can still be performed with the effective multipole moments retrieved from numerical simulations. In Figure 2(b), we have plotted the former, given by

$$r = \frac{C}{(2a+s)^2}\left(\tilde{p}_x - \frac{1}{c}\tilde{m}_y + \frac{ik}{6}\tilde{Q}_{xz} - \frac{ik}{2c}\tilde{M}_{yz}\right). \tag{7}$$

Eq.(7) has an explicit inverse quadratic dependence with the wall-to-wall separations $s$, which ultimately determines the increase in reflection for small distances between the meta-atoms. This behavior can be clearly appreciated in the absolute values of the multipolar contributions shown in Figure 2(b). We emphasize that the quadratic dependence is common to all subwavelength square arrays, not only the ones studied here, and is due to an averaging of the electric field over $S_l$ (see Eq.(S2) in the Supplementary Material S1). A unique trait of HA metasurfaces lies in their ability to minimize this effect due to the small values of the multipolar contributions to the far-field that enter the numerator of Eq.(7), while simultaneously featuring strongly confined near fields. Most importantly, the effective multipoles are almost identical to the isolated particle (refer to Figure S4 in the Supplementary Material S7), implying that even for very small separations they are not affected by the scattered fields of the neighbors. Thus, the slight increase in reflection is not due to coupling, which remains *negligible*.

We illustrate this aspect further by calculating the near fields for two metasurfaces with different period, the most compact one corresponding to an extreme case when the smallest spherical surfaces enclosing two neighboring meta-atoms intersect each other [Figure 2(c)]. For a conventional nanoantenna, the latter would usually lead to strong near field coupling [49]. In stark contrast with this initial intuition, an inspection of the fields within the high-index regions of the two metasurfaces [Figures 2(c)-(d)], reveals an identical picture.

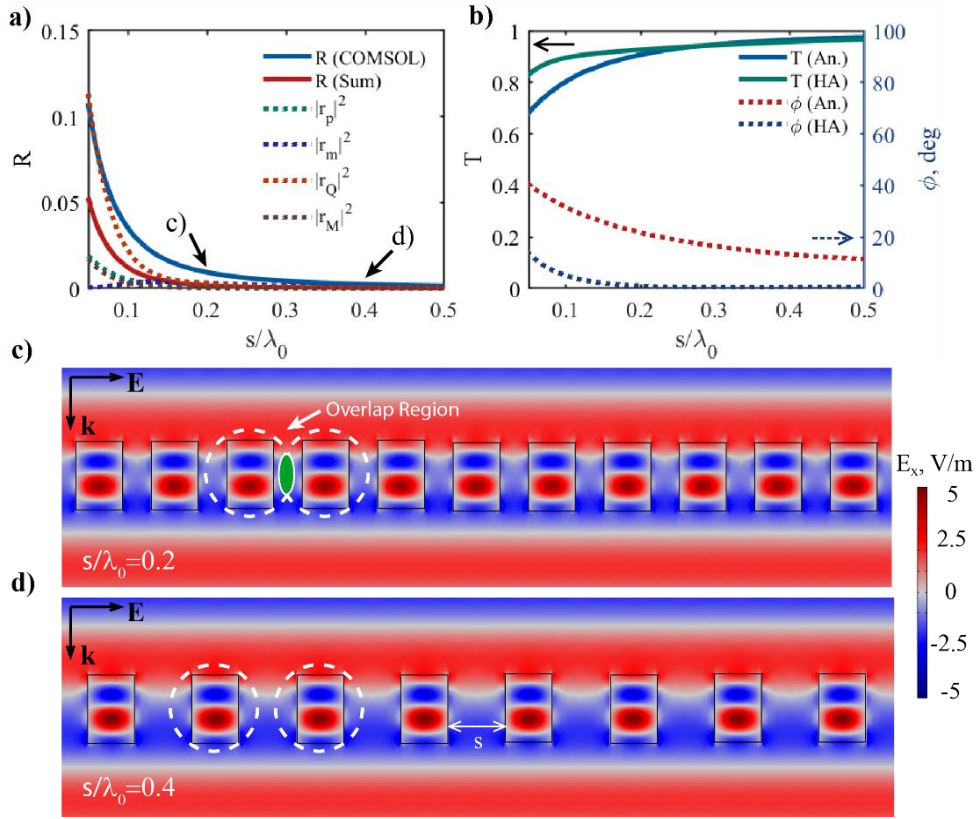

**Figure 2**. Optical response of the HA metasurface with particle separation $s$ (the geometrical parameters are the same as in Figure 1). Evolution of reflection (a), and transmission (b) at $\lambda_0$. (c)-(d) x-component of the total electric field for $s/\lambda_0 = 0.2 - 0.4$. The dashed circles depict a section of the smallest spherical surface enclosing the scatterer. For $s/\lambda_0 = 0.2$ they overlap, demonstrating that our metasurface retains the same phase and transmission even in the extreme case when the 'equivalent' spheres would intersect. In both scenarios, the fields within the meta-atoms remain virtually identical, and the incident field is completely transmitted.

Motivated by these initial findings, we now aim at designing a 'map' relating a specific transmitted phase with some geometrical parameter of the meta-atom. Keeping in mind a future practical realization, we choose radius $a$ as our degree of freedom, since radial deformations can be implemented with well-established fabrication techniques, e.g. electron beam lithography followed with reactive ion etching [13]. To this end, we calculate the transmission and phase for the HA metasurface with small separations $s = 150$ nm for different radii and wavelengths [Figure 3(a,b)]. In contrast to a sphere, the behavior with $a$ is not identical to the one with $\lambda$. This owes to the fact that the HA appears due to the overlap of Mie and Fabry-Perot modes having different dependence with $a$ [30]. Nevertheless, we can set a wavelength of operation and observe a well-defined phase variation for a collection of radii within the HA-mediated transparency window [Figure 3(c)], which we refer to as an 'Abacus' [13,42]. We have calculated two Abacus for different $s$, which yield almost the same results, thus once more showcasing the ability of HA nanoparticles to act as individual meta-atoms.

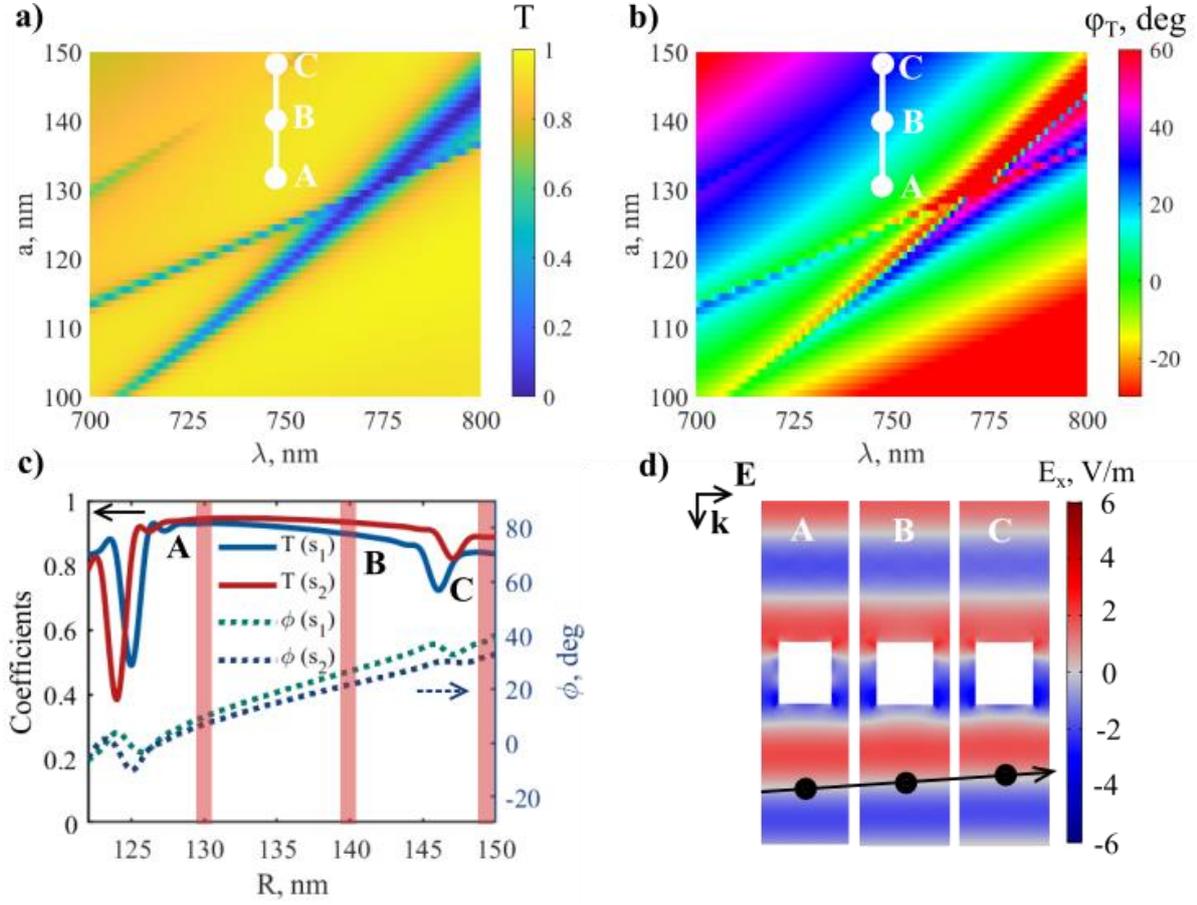

**Figure 3.** HA metasurfaces enabling phase control with full transmission. (The height of the cylinders is kept constant at 370 nm) (a) Transmission as a function of wavelength and radius of the meta-atom, calculated at a fixed, (arbitrarily chosen) wall-to-wall distance between cylinders $s = 150$ nm. The chosen range corresponds to the maximum available interval of radii featuring suppressed reflection ($\lambda = 750$ nm). (b) Phase of the transmitted wave for the same parameters as in (a). (c) The 'Abacus' of the proposed metalattice in the 0th diffractive order, calculated for two ultra-compact arrays with $s = 100, 150$ nm at $\lambda = 750$ nm. (d) From left to right: the x component of the electric field in three metasurfaces constituted, respectively, of meta-atoms with radius $a = 130, 140, 150$ nm, and transmitted phases $\phi_T = 10, 30, 40$ deg, selected from (b) (indicated with A,B,C at $\lambda = 750$ nm).

The Abacus allows us to select a meta-atom yielding a specific phase depending entirely on $a$. This behavior is shown in Figure 3(d), where we have plotted the **x** component of the total electric field for the points A,B,C indicated in Figure 3(c). Indeed, the transmitted wave modifies its delay with the incident field when transitioning from A to C in parameter space.

Summarizing the results, we have first designed a new kind of transparent metasurface based on the low scattering properties of the HA, and rigorously analyzed its working principle through a multipolar decomposition of reflection and transmission. The HA opens a transmission band with the possibility to alter the phase of the transmitted wave. In the case studied, the phase change occurs mainly due to an increase in the contribution of the electric quadrupole moment, whose amplitude is strongly minimized due to the electric quadrupole anapole. In accordance with our initial predictions, we have demonstrated numerically that our design truly suppresses mutual interaction between neighbors and elaborated an 'Abacus' encoding a specific phase delay to HA nanocylinders of a given radius. In the next section, we will take a step further and unveil the true potential of our design for the fabrication of phase-changing metasurfaces exhibiting disorder.

## 3.2 Disordered lattices

Despite the large bulk of literature dedicated to the study of periodic lattices, their fabrication requires the implementation of very precise nanolithography techniques. Thus, to a certain degree, all realistic metasurfaces exhibit disorder. The presence of disorder can drastically alter the optical response of the array, making the design of applications a challenging task[50] . Essentially, from Eq. (1) two types of randomness can be introduced [51]: (i) size disorder directly affecting the polarizability matrix $\mathcal{A}$ , or (ii) positional disorder (PD), which modifies the components of the coupling matrix $\mathcal{S}$ . Since we are only interested in the lattice effects, in what follows we consider solely (ii).

In conventional metasurface designs, the influence of PD is essentially dependent on the mutual interaction between the meta-atoms. For instance, the collective lattice resonances of arrays of Si nanospheres were shown to be suppressed under the influence of certain types of lattice perturbations [52]. Importantly, due to their strong sensitivity to the spacing among neighboring nanodisks [53], Huygens metasurfaces are strongly affected by disorder and exhibit spontaneous transitions at critical values of PD, featuring abrupt variations of the transmitted phase [54]. Here, once again, we encounter an important advantage offered by the HA meta-atom; indeed, disorder effects must also be strongly suppressed due to the negligible polarizabilities entering $\mathcal{A}$ .

To demonstrate the immunity to PD, we perform two sets of numerical experiments with the results displayed in Figure 4 (a) - (e). In Figure 4(a)-(b), we introduce an in-plane PD in the metalattice, controlled by a normal distribution of the lattice period with mean $l$ , and standard deviation $\delta l$ . We numerically implemented the disorder when constructing a finite $4 \times 4$ lattice and then imposed periodic boundary conditions, as depicted in Figure 4(c), which also shows the displacement of each HA meta-atom with respect to the ideal array. Similarly, in Figure 4(d)-(e), the PD is induced out-of-plane, along the z-axis, as illustrated in Figure 4(f). Out-of-plane PD can be associated, e.g., with surface roughness.

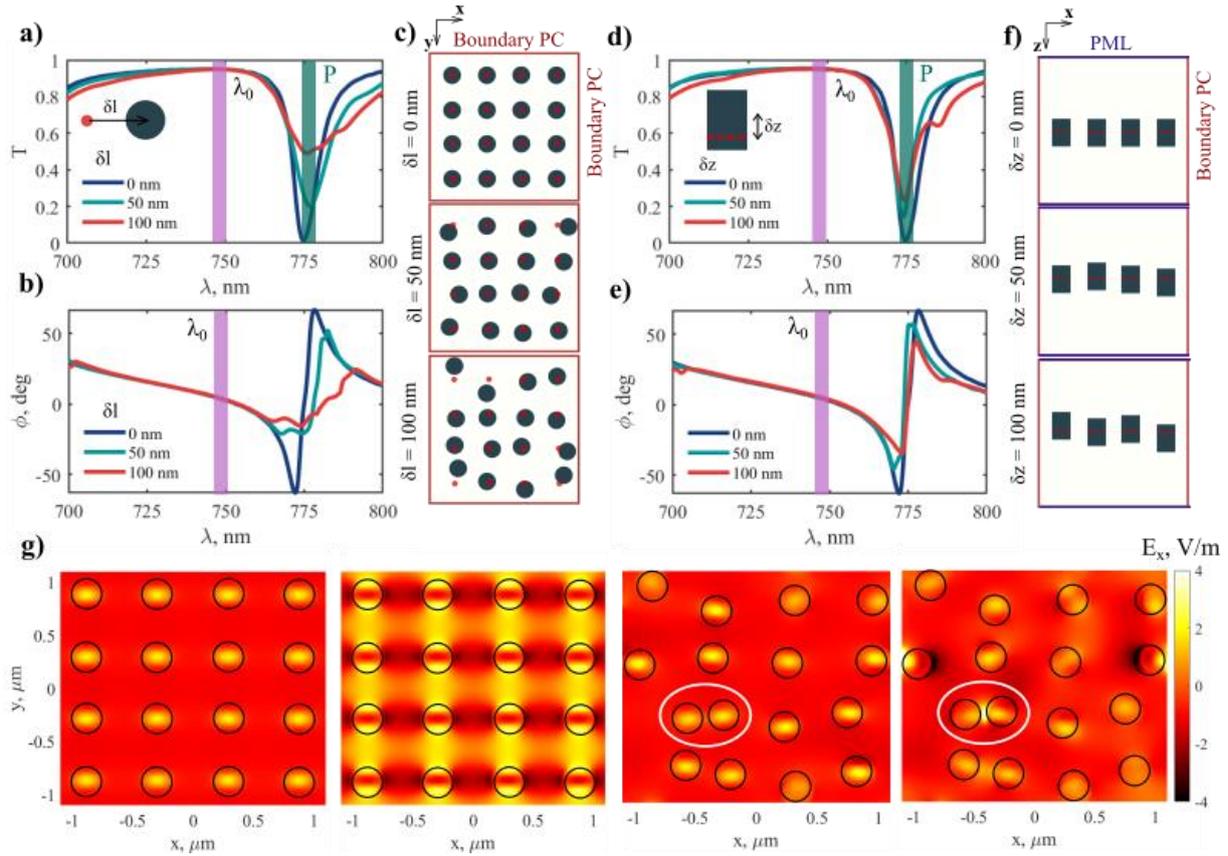

**Figure 4:** Numerical experiments demonstrating the resilience of HA metasurfaces to PD, paving the way towards disorder-immune metadevices (geometrical parameters as in Figure 1). (a)-(b) Transmission and transmitted phase with different degrees of in-plane PD with standard deviation $\delta l$, schematically depicted in (c), where the red dots indicate the ideal periodic array. (d)-(e) Transmission of the array with different degrees of out-of-plane PD, with standard deviation $\delta z$, schematically depicted in (f). In the two cases studied, both transmission and phase in the HA regime are well preserved and remain almost invariant at $\lambda_0$. (g) Calculated distributions of the x-component of the electric field at $\lambda_0$ and point P indicated in (a) and (d). All fields are normalized to the same colorbar. From left to right, the first two field maps correspond to the periodic array at $\lambda_0$ and point P, respectively. The other two correspond to the same points for a PD given by $\delta l = 100$ nm. The white circles highlight two closely packed nanorods exhibiting a markedly different behavior; despite their proximity, at $\lambda_0$ no interparticle hotspot can be appreciated while the dimer system at point P can be clearly seen to display strong near field interactions.

Remarkably, in Figure 4(a)-(f), only small variations of the transmission spectrum and the transmitted phase can be observed in the wavelength range affected by the HA, demonstrating its striking resilience to very large deviations from the ideal periodic lattice. In stark contrast, a strong dependence on PD at the wavelengths 'unprotected' by the HA can be clearly distinguished, particularly prominent at the dip of the transmission spectra, [point P in Figure 4(a) and Figure 4(d)], where a large change in the transmission amplitude takes place. These behaviors are also confirmed in the near field distributions at $\lambda_0$ and at the dip of the transmission spectrum. On the one hand, at point P, as a consequence of strong scattering, a drastic change in the fields can be appreciated when transitioning from the periodic to the disordered array, both outside and within the nanoparticles. In the rightmost panel of Figure 4(g) (disordered array illuminated at point P), strong evanescent fields appear when the particles are close to each other. On the other hand, the near fields of the disordered HA metasurfaces remain visually unaltered, even for particles that would usually exhibit large evanescent fields, such as the pair of nanorods highlighted on the second rightmost panel of Figure 4(g). In addition, it is worth mentioning that the proposed metasurface benefits from the strong field concentration characteristic of anapole-like regimes. Therefore, disordered HA metasurfaces constitute a flexible platform to enhance light-matter interactions at the nanoscale in a simple and straightforward fashion, without the need of complex optimization techniques nor requiring a careful arrangement of the meta-atoms.

From the results in this section, we can conclude that both phase and transmission of HA meta-atoms are 'protected' against disorder. Thus, HA-based metasurfaces might not require periodicity in order to implement a varying spatial phase profile, offering exciting perspectives for the realization of applications.

### 3.3 Influence of a dielectric substrate

A practical implementation of metasurfaces would unavoidably require the presence of a substrate. The latter can play a non-negligible role in the optical response and introduces magnetoelectric coupling [55]. In stark contrast with conventional resonances, the HA is remarkably robust when deposited over a substrate [30]. In the studied nanocylinder, the HA (differently from conventional anapoles or Huygens sources) is effectively attained through the overlap of resonant 'Mie-like' and 'Fabry-Perot-like' modes. The first can be associated to standing waves originating between the lateral walls of the resonator cavity, while the second is mainly formed from standing waves between the top and bottom walls. Therefore, variations in the substrate reflectivity affect mainly the amplitude of Fabry-Perot modes, but the Mie modes remain almost unaltered [30]. With a variation of $n_{sub}$, the HA gradually transforms to a conventional electric dipole anapole, still retaining a strong scattering reduction.

HA metasurfaces are expected to present similar features (Figure 5). In Figure 5(a) we have calculated transmission, reflection and absorption at the HA regime for our metasurface deposited over a series of hypothetical substrates with index ranging from $n_{sub} = 1,...,2$. We observe a progressive narrowing of the transmission band mainly resulting from a redshift of high order Bloch modes. Importantly, full transmission is preserved in the vicinity of $\lambda_0$, with a decrease of only 10% for $n_{sub} = 2$. In Figure 5(b) we have calculated the $E_x$ component of the total field, which clearly shows how the wave at $\lambda_0$ is almost fully transmitted by the metasurface even for $n_{sub} = 2$.

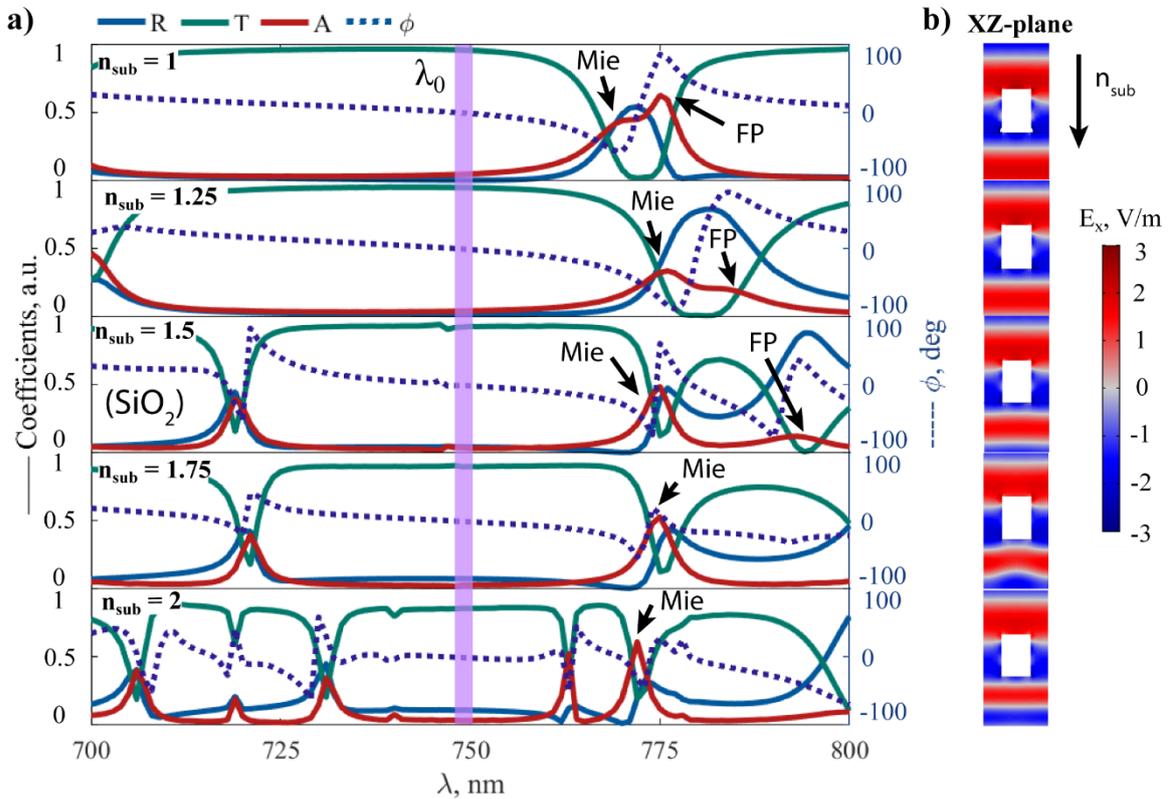

**Figure 5.** (a) Optical response of a HA metasurface with inter-particle separation s = 300 nm, and geometrical parameters as in Figure 1, deposited on hypothetical substrates with increasing refractive index. The arrows indicate the evolution of the Mie-like and Fabry-Perot (FP) modes responsible for the HA. (b) x-component of the electric field at $\lambda_0$ for the cases considered in (a).

Unlike in the single particle case, the Mie and Fabry Perot modes are coupled. The drop in transmission and the change in the transmitted phase are linked to the evolution of the Fabry-Perot mode; as indicated in Figure 5(a), the substrate index influences strongly the width and spectral position of the Fabry-Perot resonance, that rapidly redshifts while the Mie-mode remains almost unaltered. As a result, the HA starts degenerating into a conventional anapole. However, the transmission band induced by the HA is very resilient to changes in the underlying substrate. Importantly, the results unambiguously show that the metasurface can be directly deposited over conventional silica (SiO2) without further design steps and display full transmission as well as enabling phase control. In this case, in exchange for a small reduction of the transmission band, the available range for phase tunability is doubled with respect to free space to approximately 80 degrees, suggesting there is still a large room for improvement of the effect.

# 4 Phase modulation of fs pulses with a disordered HA array

Next, we discuss the possibility to modulate the phase of an ultrafast gaussian pulse in transmission mode, making use of a disordered HA array on top of a silica substrate. To illustrate the robustness of our system, we choose a strong in-plane disorder ($\delta l = 50\,\text{nm}$), corresponding to 20% the separation between nanorods ($s = 300\,\text{nm}$). We then perform Finite-Difference Time Domain simulations (FDTD) of an incoming x-polarized gaussian pulse with 400 fs duration. The concept is depicted in Figure 6(a); the radius of the nanorods is chosen in accordance to an Abacus similar to the one in Figure 3(c), but calculated with the periodic array placed on top of the glass substrate (refer to Supplementary Material S8). With a progressive increase in $a$, the phase of the output pulse can be controllably tuned, as shown in Figure 6(b), where we have plotted the maxima of the output and input pulses. We have chosen $a$ to vary between 125 and 135 nm, to be well in the range covered by the HA (Figure S5). For comparison, in Figure 6(b) we have also added the results for the periodic array.

As could be expected from our calculations in Figure 4, the phase imprinted by the periodic and the disordered metasurfaces are virtually identical, once more demonstrating that HA nanoparticles indeed operate in an almost independent fashion from their neighbors. Consequently, transmission is kept higher than 85 %, as confirmed by the full temporal profiles displayed in Figure 6(c)-(d) (dissipative losses were not neglected in the simulations). Alternatively, although generally less practical, the device could also be controlled in wavelength and designed according to the results in Figure 5(a).

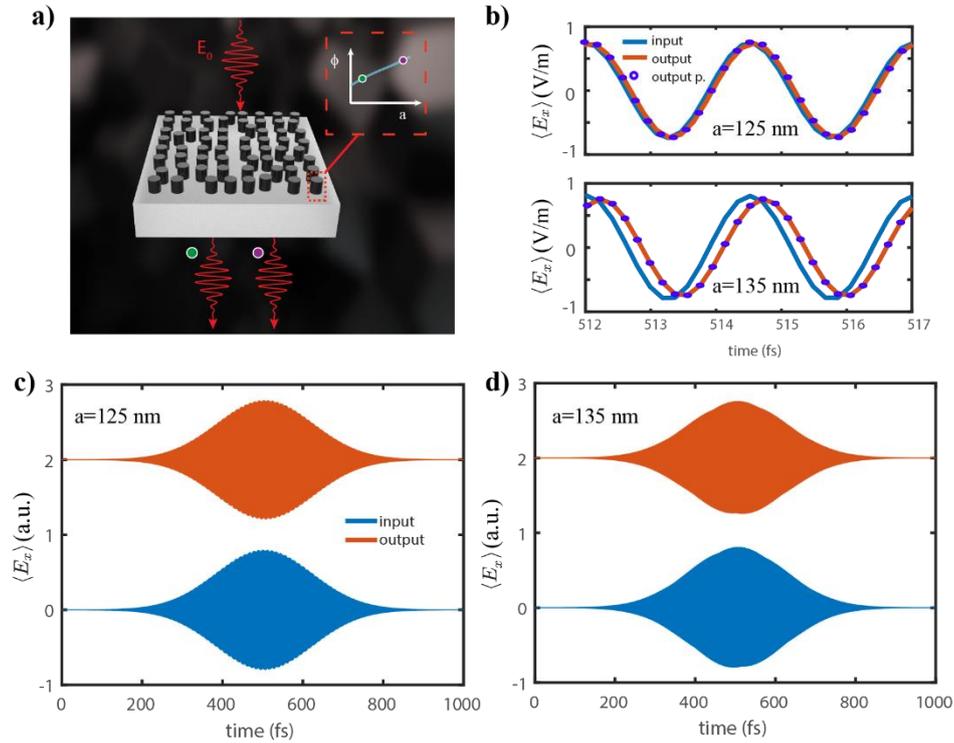

**Figure 6.** Manipulating the phase of an x-polarized fs Gaussian pulse with a disordered HA metasurface on a semi-infinite glass substrate. (a) Artistic representation of the concept, depicting the impinging and outgoing pulses from the disordered HA metasurface with tailored nanoparticle sizes, and a schematic of the Abacus. (b) Transparency and advanced phase of the output pulse at their maxima, for $a = 125, 135$ nm (top and bottom panels, respectively) in the disordered lattice, in perfect agreement with the periodic arrangement (output p.). The height of the nanorods is kept constant at 370 nm. (c)-(d) Temporal profiles of the incoming (input) and outgoing pulses (output), demonstrating full transmission for the two cases studied. In panels (b)-(d) the operator '$\langle \ \rangle$' denotes averaging over the transverse cross section of a single unit cell.

Summarizing, we have numerically implemented an ultrafast phase modulator based on a disordered HA array, relying solely on the knowledge of the periodic lattice. The results suggest the possibility to flexibly design metadevices without the need of time-consuming optimization steps, reaching very high resolution. It is worth noticing that even better performances could be obtained in the mid-IR range, since our setup is easily scalable, and could be introduced in a plethora of active and passive light spatial modulation schemes.

# 5 Conclusion

We have proposed and investigated in detail *novel transparent HA metasurfaces* based on the unusual properties of the recently observed HA regime. Unlike recent designs exploiting, e.g., the transverse Kerker effect [29], our metasurfaces achieve not only near unity efficiency in transmission, but also allow to vary the phase of the transmitted wave. Most importantly, they display negligible inter-particle coupling, overcoming a critical bottleneck of conventional Huygens sources. We have demonstrated how the latter leads to a number of technological advantages, including the possibility to fabricate ultra-compact arrays retaining the single particle response, the preservation of the transmission window when deposited on a wide range of dielectric substrates, and a robust protection against large PD, significantly relaxing the technical requirements of the manufacturing process. As a proof-of-concept, we have numerically demonstrated an ultrafast phase modulator operating in transmission with a disordered and a periodic HA array, showing perfect agreement between the two structures. We emphasize that the HA regime is not limited by strict design constraints and can be easily implemented in (but not only) an amorphous silicon nanorod in the visible range under normally incident plane wave illumination. From the above, we can conclude that HA nanoparticles, in contrast with Huygens disks, truly approach the ideal of a 'meta-atom', i.e. a subwavelength unit cell with a unique phase imprint. In the future, we believe that clever designs of HA-based metasurfaces could compete with their more established counterparts, paving the way towards new devices in meta-optics.

## Acknowledgments

A.V.K and A.C.V contributed equally to this work. The authors gratefully acknowledge the support of the RFBR Grant 20-52-00031 for the multipolar decompositions. The transient calculations for the disordered metasurfaces have been partially supported by the RSF Grant 21-12-00151.

# Supplementary Material for "Transparent Hybrid Anapole Metasurfaces with negligible Electromagnetic Coupling for Phase Engineering"

Alexey V. Kuznetsov[1], Adrià Canós Valero[1], Mikhail Tarkhov[4], Vjaceslavs Bobrovs[2], Dmitrii Redka[2,3], and Alexander S. Shalin[1,2,5]

[1]ITMO University, Kronverksky prospect 49, 197101, St. Petersburg, Russia

[2]Riga Technical University, Institute of Telecommunications, Latvia, Riga, Azenes street 12, post code 1048

[3]Electrotechnical University "LETI" (ETU), 5 Prof. Popova Street, Saint Petersburg 197376, Russia

[4]Institute of Nanotechnology of Microelectronics of the Russian Academy of Sciences (INME RAS), Moscow, Nagatinskaya street, house 16A, building 11

[5]Kotel'nikov Institute of Radio Engineering and Electronics of Russian Academy of Sciences (Ulyanovsk branch), Goncharova Str.48, Ulyanovsk, Russia, 432000

## S1. Multipolar decomposition of reflection and transmission from a subwavelength periodic array

In Ref. [5] the exact multipolar decomposition of reflection and transmission was derived making use of multiple scattering theory. While providing with analyticity, the approach is only rigorously valid if the smallest spheres enclosing the scatterer do not intersect with each other. We now put forward a simple derivation based on more general principles that demonstrates the correctness of Eq.(6) in the main text even when multiple scattering theory cannot be applied to the system in question.

Consider an infinite periodic array located in the x-y plane, and a point located in the z-axis placed in the far field. The electric field at this point $\mathbf{r} = (0 \quad 0 \quad z)^T$ would correspond to the superposition of the scattered fields from all the particles in the array and the incident field $\mathbf{E}_0(\mathbf{r})$:

$$\mathbf{E}(\mathbf{r}) = \mathbf{E}_0(\mathbf{r}) + \sum \mathbf{E}_l(\mathbf{r}). \tag{S1}$$

Clearly the scattered field $\mathbf{E}_{sca}(\mathbf{r}) = \sum \mathbf{E}_l(\mathbf{r})$. Now, if the unit cell is subwavelength, we can replace the sum by an integral over the transverse area of the unit cell[6]:

$$\mathbf{E}_{sca}(\mathbf{r}) \approx \frac{1}{S_l} \int \mathbf{E}_l(\mathbf{r}) dA. \tag{S2}$$

In the far field, the multipole decomposition of $\mathbf{E}_l(\mathbf{r})$ is given by (up to the quadrupole terms):

$$\mathbf{E}_l(\mathbf{r}) \sim \frac{k_0^2}{4\pi\varepsilon_0} \frac{e^{ikr}}{r} \Bigg( [\mathbf{n} \times (\tilde{\mathbf{p}} \times \mathbf{n})] + \frac{1}{v_d}[\tilde{\mathbf{m}} \times \mathbf{n}] + \\ \frac{ik_d}{6}[\mathbf{n} \times (\mathbf{n} \times \tilde{Q}\mathbf{n})] + \frac{ik_d}{2v_d}[\mathbf{n} \times (\tilde{M}\mathbf{n})] \Bigg), \tag{S3}$$

where $\mathbf{n} = \mathbf{r}/r$. In our case $\mathbf{n}_{\pm} = (0 \quad 0 \quad \pm 1)^T$. Substituting Eq.(S3) into Eq.(S2) leads to an expression for $\mathbf{E}_{sca}(\mathbf{r})$, which can then be directly used to evaluate the reflection and transmission coefficients as

$$r = [\mathbf{E}_{sca} / \mathbf{E}_0]_{\mathbf{n}_-}, \tag{S4}$$

$$t = 1 + \left[\mathbf{E}_{sca}/\mathbf{E}_0\right]_{\mathbf{n}_+}. \tag{S5}$$

For an x-polarized plane wave propagating in the $-z$ direction, the forward normal vector is $\mathbf{n}_+ = [0,0,-1]$, and the backward vector is $\mathbf{n}_- = [0,0,1]$. The angular distribution of the scattered field in any of the two directions is then:

$$E_{x,\pm}^{sca}(n_\pm) \sim \left( \tilde{p}_x n_\pm^2 + \frac{1}{v_d}\tilde{m}_y n_\pm - \frac{ik_d}{6}\tilde{Q}_{xz}n_\pm^3 \right. \\ \left. - \frac{ik_d}{2v_d}\tilde{M}_{yz}n_\pm^2 - \frac{k_d^2}{6}\tilde{O}_{xzz}n_\pm^4 \right) \tag{S6}$$

The reflection coefficient is[7]

$$r = \frac{ik_0}{2S_l \varepsilon_0 E_0(z)} E_{x,-}^{sca}, \tag{S7}$$

while the transmission coefficient can be written as[7]

$$t = 1 + \frac{ik_0}{2S_l \varepsilon_0 E_0(z)} E_{x,+}^{sca}. \tag{S8}$$

Eqs. (S7)-(S8) are equivalent to Eqs. (6)-(7) in the main text. It is now clear that the only condition necessary for the expressions to be valid is that the *unit cell must be subwavelength*, in order to apply Eq.(S2).

## S2. Multiple scattering theory

In this section we briefly summarize well-known results regarding multiple scattering theory. For simplicity in the treatment, we consider scatterers with only electric and magnetic dipolar response, and refer the interested reader to the detailed series of works [1,2] for expressions including the effective quadrupole terms.

In this simple scenario, we analyze an array of subwavelength nanoparticles which are also small with respect to their mutual separation. Then, the effective electric (magnetic) dipole moments are induced by the instantaneous electric (magnetic) fields 'seen' by each constituent, namely the incident field and the sum of all fields scattered by the other particles in the array:

$$\tilde{\mathbf{p}}_l = \alpha_p \left[ \mathbf{E}_0(\mathbf{r}_l) + \sum_{n \neq l} \mathcal{G}^0(\mathbf{R}_{ln})\tilde{\mathbf{p}}_n + \frac{ik_0}{c\varepsilon_0}\mathbf{g}(\mathbf{R}_{ln})\times\tilde{\mathbf{m}}_n \right], \tag{S9}$$

$$\tilde{\mathbf{m}}_l = \alpha_m \mathbf{H}_0(\mathbf{r}_l) + \alpha_m \sum_{n \neq l} \left[ \varepsilon_0 \mathcal{G}^0(\mathbf{R}_{ln})\tilde{\mathbf{m}}_n - ik_0 c \mathbf{g}(\mathbf{R}_{ln})\times\tilde{\mathbf{p}}_n \right]. \tag{S10}$$

In the previous $\alpha_p$, $\alpha_m$ are the particle polarizabilities (inherent to a single scatterer), $\mathbf{R}_{ln} = \mathbf{r}_l - \mathbf{r}_n$, $\mathcal{G}^0(\mathbf{R}_{ln})$ is the free space electromagnetic propagator evaluated at $\mathbf{R}_{ln}$, and given by

$$\mathcal{G}^0(\mathbf{r}) = (k^2 + \nabla\nabla)\frac{e^{ikr}}{r}. \tag{S11}$$

The vector $\mathbf{g}(\mathbf{R}_{ln})$ is introduced for convenience:

$$\mathbf{g}(\mathbf{R}_{ln}) = \frac{e^{ik_0 R_{ln}}}{4\pi R_{ln}}\left(\frac{ik_0}{R_{ln}} - \frac{1}{R_{ln}^2}\right)\mathbf{R}_{ln}. \tag{S12}$$

Eqs. (S9)-(S10) form a linear system of equations for all the effective multipoles. Under normally incident x-polarized plane wave, assuming all the particles are distributed in the x-y plane, only x-polarized electric dipoles and y-polarized magnetic dipoles can be excited, and the electric and magnetic moments are not coupled (the third term in the rhs of Eqs. (S9)-(S10) vanishes). Then we can group all multipoles in a vector $\tilde{\mathbf{b}} = (\tilde{p}_1, \tilde{p}_2, ..., \tilde{p}_N, \tilde{m}_1, ..., \tilde{m}_N)^T$, and write a more compact formula:

$$[\mathcal{U} - \mathcal{A}\mathcal{S}]\tilde{\mathbf{b}} = \mathbf{b},  \tag{S13}$$

where the vector $\mathbf{b}$ contains the multipole moments induced solely by the incident field, $\mathcal{U}$ is the identity matrix, $\mathcal{A}$ is a diagonal matrix containing the single particle polarizabilities, and $\mathcal{S}$ is the coupling matrix, the elements of which in this example have the form

$$\mathcal{S}^p_{ln} = (1 - \delta_{ln}) \mathcal{G}^0_{xx}(\mathbf{R}_{ln}), \tag{S14}$$

$$\mathcal{S}^m_{ln} = (1 - \delta_{ln}) \varepsilon_0 \mathcal{G}^0_{yy}(\mathbf{R}_{ln}) \tag{S15}$$

for the electric and magnetic parts of $\tilde{\mathbf{b}}$, respectively.

## S3. Cartesian Multipole Expansion

The exact Cartesian expressions of the electromagnetic multipole moments up to the quadrupoles were recently derived in Ref.[3], and are given by the following integrals over the resonator volume:

$$\mathbf{p} = \int \mathbf{P} j_0(kr) d^3\mathbf{r} + \frac{k^2}{10} \int \left\{ [\mathbf{r} \cdot \mathbf{P}]\mathbf{r} - \frac{1}{3} r^2 \mathbf{P} \right\} \frac{15 j_2(kr)}{(kr)^2} d^3\mathbf{r}, \tag{S16}$$

$$\mathbf{m} = -\frac{3i\omega}{2} \int (\mathbf{r} \times \mathbf{P}) \frac{j_1(kr)}{kr} d^3\mathbf{r}, \tag{S17}$$

$$Q = \int \{3(\mathbf{r} \otimes \mathbf{P} + \mathbf{P} \otimes \mathbf{r}) - 2[\mathbf{r} \cdot \mathbf{P}]\mathcal{U}\} \times \frac{3 j_1(kr)}{kr} d^3\mathbf{r} +$$
$$6k^2 \int \{5\mathbf{r} \otimes \mathbf{r}[\mathbf{r} \cdot \mathbf{P}] - (\mathbf{r} \otimes \mathbf{P} + \mathbf{P} \otimes \mathbf{r})r^2 - r^2[\mathbf{r} \cdot \mathbf{P}]\mathcal{U}\} \frac{j_3(kr)}{(kr)^3} d^3\mathbf{r}, \tag{S18}$$

$$M = \frac{\omega}{3i} \int \{[\mathbf{r} \times \mathbf{P}] \otimes \mathbf{r} + \mathbf{r} \otimes [\mathbf{r} \times \mathbf{P}]\} \frac{15 j_2(kr)}{(kr)^2} d^3\mathbf{r}. \tag{S19}$$

Where $\mathbf{P}$ is the induced polarization current within the scatterer, and the $j_i(kr)$ are the $i$th spherical Bessel functions of the first kind. Taking the first order terms of the Taylor series of Eqs.(S16)-(S19), one recovers the quasistatic contributions having the classical expressions:

$$p^0_\alpha = \int P_\alpha d^3\mathbf{r}, \tag{S20}$$

$$m^0_\alpha = -\frac{i\omega}{2} \int (\mathbf{r} \times \mathbf{P})_\alpha d^3\mathbf{r}, \tag{S21}$$

$$Q^0_{\alpha\beta} = 3\int \left[ \left( r_\beta P^\omega_\alpha + r_\alpha P^\omega_\beta \right) \right] d^3\mathbf{r}, \tag{S22}$$

$$M^0_{\alpha\beta} = -i\omega \int \left\{ r_\alpha (\mathbf{r} \times \mathbf{P})_\beta + r_\beta (\mathbf{r} \times \mathbf{P})_\alpha \right\} d^3\mathbf{r}. \tag{S23}$$

In Eqs. (S22)-(S23) we have omitted the diagonal terms of the tensors since they cannot be excited with conventional illumination schemes. The first toroidal terms are given by[4]

$$T_\alpha^p = \frac{k^2}{10}\int\left[-2r^2 P_\alpha +(\mathbf{r}\cdot\mathbf{P})r_\alpha\right]d^3\mathbf{r}, \tag{S24}$$

$$T_\alpha^m = \frac{i\omega k^2}{20}\int\left[(\mathbf{r}\times\mathbf{P})r^2\right]d^3\mathbf{r}, \tag{S25}$$

$$T_{\alpha\beta}^Q = \frac{k^2}{14}\int\left[4r_\alpha r_\beta(\mathbf{r}\cdot\mathbf{P})-5r^2(r_\alpha P_\beta + r_\beta P_\alpha)\right]d^3\mathbf{r}, \tag{S26}$$

$$T_{\alpha\beta}^M = \frac{i\omega k^2}{14}\int r^2\left[r_\alpha(\mathbf{r}\times\mathbf{P})_\beta + r_\beta(\mathbf{r}\times\mathbf{P})_\alpha\right]d^3\mathbf{r}. \tag{S27}$$

We bring the attention of the reader to the fact that Eqs.(S24)-(S27) have different normalization factors than the expressions introduced in Ref. [4]. This is done so that the total contribution of the current multipoles to the scattered field outside the resonator can be conveniently written as the sum of the quasistatic and the toroidal parts. For example, the scattered field of electric dipole type is proportional to $p_\alpha = p_\alpha^0 + T_\alpha^p$, where the absence of superscript indicates 'exact'. Similarly, the other 'exact' multipoles will be indicated in this fashion.

## S4. Experimental Dispersion of amorphous Si

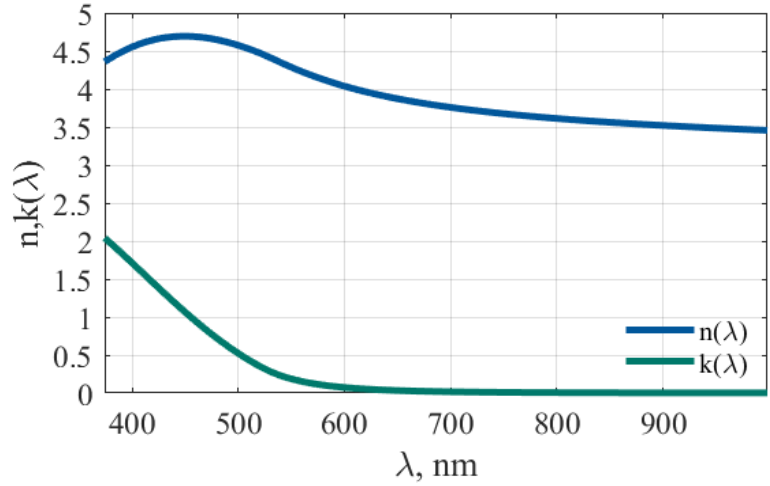

**Figure S1.** Real (n) and imaginary (k) parts of the experimentally measured refractive index of amorphous silicon (aSi) employed in our study, as a function of wavelength.

## S5. Multipole decomposition of the scattering cross section for an isolated nanoparticle

In an homogeneous environment, the contributions to the cross section of the different 'exact' multipole moments are decoupled, and can be written as (up to the quadrupole)[8]:

$$\sigma_{sca} \simeq \frac{k_0^4}{6\pi\varepsilon_0^2|\mathbf{E}_{inc}|^2}|\mathbf{p}|^2 + \frac{k_0^4\mu_0}{6\pi\varepsilon_0|\mathbf{E}_0|^2}|\mathbf{m}|^2 + \\ \frac{k_0^6}{720\pi\varepsilon_0^2|\mathbf{E}_0|^2}\sum_{\alpha\beta}|Q_{\alpha\beta}|^2 + \frac{k_0^6\mu_0}{80\pi\varepsilon_0|\mathbf{E}_0|^2}\sum_{\alpha\beta}|M_{\alpha\beta}|^2 \tag{S28}$$

We plot each contribution, together with the full-wave solution in Figure S2, demonstrating perfect agreement, as well as the appearance of a dip in every dominant multipole, leading to a Hybrid Anapole state.

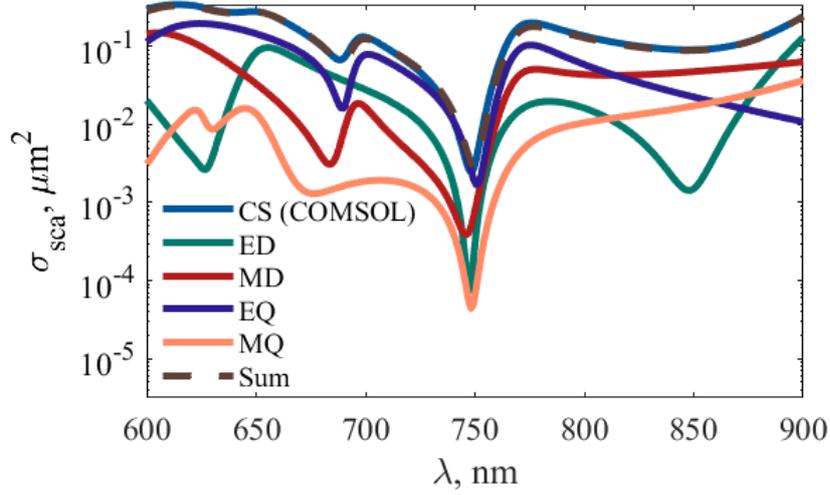

**Figure S2.** Calculated and reconstructed scattering cross section of an isolated aSi nanorod with height 370 nm and radius 125 nm. The abbreviations in the legend denote the electric dipole (ED), magnetic dipole (MD), electric quadrupole (EQ), and magnetic quadrupole (MQ) contributions. $\lambda_0$ is defined at the minima, corresponding to 750 nm.

## S6. Alternative Cartesian Multipole decomposition separately taking into account toroidal moments

Eqs.(6)-(7) in the main text can be further decomposed into the quasistatic and toroidal contributions, simply by expanding the exact expressions for the multipoles entering each term in the reflection/transmission coefficient and utilizing Eqs.(S20)-(S27). The results of this procedure are shown in Figure S3(a)-(c) for the dominant multipole moments. The results unambiguously confirm that the resonant cancellation of reflection is produced when the partial fields of the quasistatic and toroidal contributions are equal in magnitude and interfere destructively.

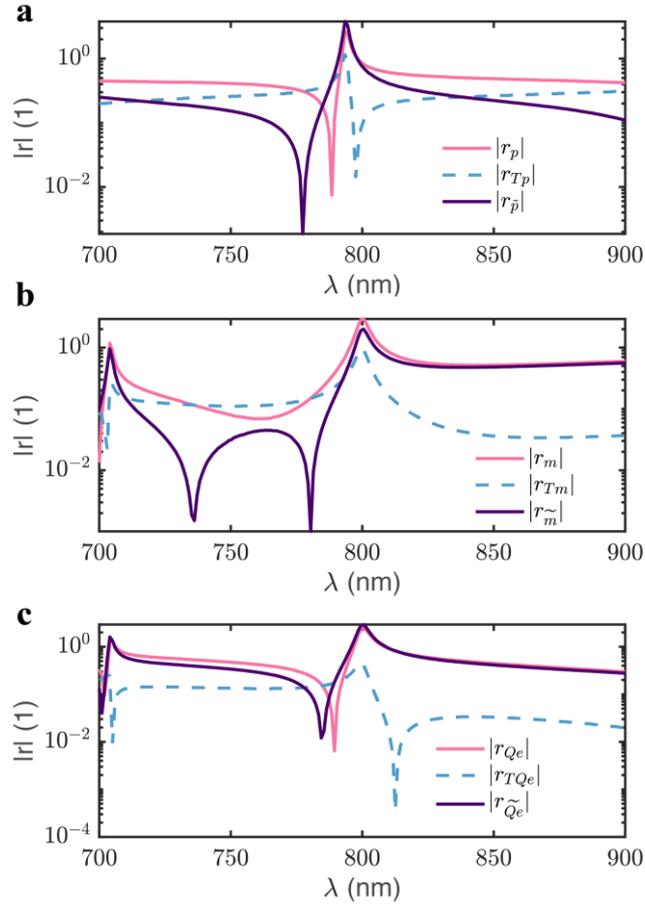

**Figure S3.** Quasistatic-toroidal multipole decomposition of the reflection coefficient for the three most relevant multipoles of an array of Hybrid Anapole nanorods like the ones studied in Ref. [9], with period 300 nm, under x-polarized plane wave illumination. (a) Contribution to reflection from the quasistatic electric dipole moment, the toroidal dipole moment, and the sum of both. (b)-(c) Same as (a), but for the magnetic dipole and the electric quadrupole moments.

## S7. Evolution of the multipole moments with interparticle separation

In Figure S4 we have plotted the effective multipole moments of the array as a function of the normalized wall-to-wall separation at $\lambda_0 = 750$ nm. The multipoles are almost constant until very small spacings, less than 6% the incident wavelength, thus confirming that coupling effects are in general negligible.

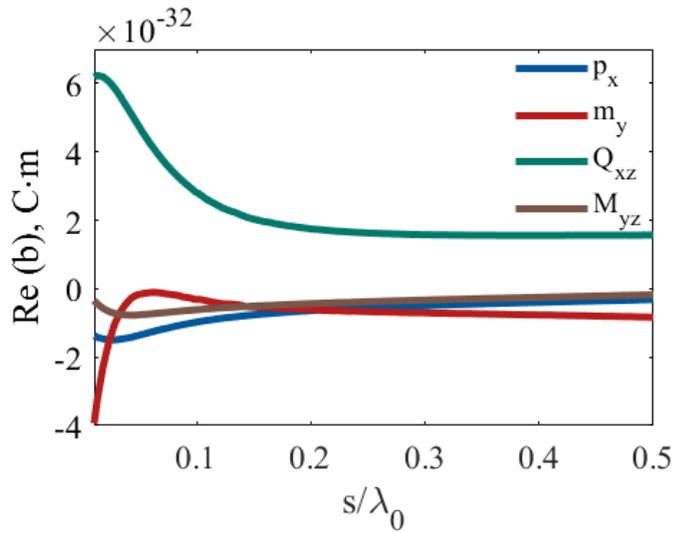

**Figure S4.** Real parts of the effective multipole moments in a unit cell as a function of normalized wall-to-wall separation $s/\lambda_0$, for a fixed wavelength of $\lambda_0 = 750$ nm. The notation follows the one used in the main text. It can be clearly seen that only very small distances result in a modification of their values, as explained in section 3 of the main manuscript.

# S8. Geometrical Abacus of the Hybrid Anapole meta-atom deposited on a glass substrate

Figure S5 shows the results of the Abacus (look-up table), when the Hybrid Anapole meta-atom is deposited on a glass substrate. The table was taken as a reference in order to control the phase of a gaussian pulse in section 4 of the main text. We can modulate the phase in a range of 80 degrees, almost doubling the original design in vacuum. The price paid is a reduction in the range of available radii, which can be tuned from 120-140 nm preserving unity transmission and negligible inter-particle coupling.

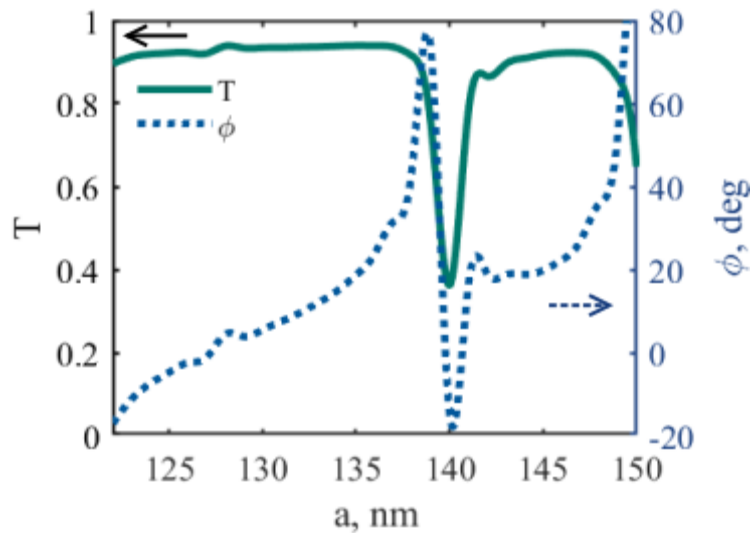

**Figure S5.** Abacus of the Hybrid Anapole meta-atom when placed on top of a glass substrate ($n_{sub} = 1.5$), for $s = 300$ nm. The plot displays transmission (left y axis), and transmitted phase (right y axis) as a function of the nanorod radius $a$.